\documentclass{llncs}

\usepackage{marvosym}
\usepackage{graphicx}
\usepackage{multirow}
\usepackage{xcolor}
\usepackage{subcaption}

\begin{document}
\title{RT-DNAS: Real-time Constrained Differentiable Neural Architecture Search for 3D Cardiac \\Cine MRI Segmentation}

\author{Qing Lu\inst{1}
\and
Xiaowei Xu\inst{2}
\and
Shunjie Dong\inst{3}
\and
Cong Hao\inst{4}
\and
Lei Yang\inst{5}
\and
Cheng Zhuo\inst{3}$^{(\textrm{\Letter})}$
\and
Yiyu Shi\inst{1}$^{(\textrm{\Letter})}$
}

\institute{
University of Notre Dame, Notre Dame, IN, USA \\
\email{\{qlu2, yshi4\}@nd.edu} \and 
Guangdong Provincial People's Hospital, Guangzhou, China \\
\email{xxu8@nd.edu} \and
Zhejiang University, Hangzhou, China \\
\email{\{sj\_dong@zju.edu.cn, czhuo@zju.edu.cn\}} \and
Georgia Institute of Technolog, Atlanta, GA, USA \\
\email{callie.hao@gatech.edu} \and
University of New Mexico, Albuquerque, NM, USA \\
\email{leiyang@unm.edu}
}
\authorrunning{Q. Lu et al.}

\maketitle              
\begin{abstract}
Accurately segmenting temporal frames of cine magnetic resonance imaging (MRI) is a crucial step in various real-time MRI guided cardiac interventions. To achieve fast and accurate visual assistance, there are strict requirements on the maximum latency and minimum throughput of the segmentation
framework. State-of-the-art neural networks on this task are mostly hand-crafted to satisfy these constraints while achieving high accuracy. On the other hand, while existing literature have demonstrated the power of neural architecture search (NAS) in automatically identifying the best neural architectures for various medical applications, they are mostly guided by accuracy, sometimes with computation complexity, and the importance of real-time constraints are overlooked. A major challenge is that such constraints are non-differentiable and are thus not compatible with the widely used differentiable NAS frameworks. In this paper, we present a strategy that directly handles real-time  constraints in a differentiable NAS framework named RT-DNAS. Experiments on extended 2017 MICCAI ACDC dataset show that compared with state-of-the-art manually and automatically designed architectures, RT-DNAS is able to identify ones with better accuracy while satisfying the real-time constraints. 

\end{abstract}

\section{Introduction}
The rapid advance in real-time cine Magnetic Resonance Imaging (MRI) has been deployed for visual assistance in various cardiac interventions, such as aortic valve replacement \cite{mcveigh2006real}, 
cardiac electroanatomic mapping and ablation \cite{radau2011vurtigo}, 
electrophysiology for atrial arrhythmias \cite{vergara2011real}, 
intracardiac catheter navigation \cite{gaspar2014three}, and
myocardial chemoablation \cite{rogers2016transcatheter}. In these applications, the cine MRI needs to be segmented on-the-fly, for which deep neural networks (DNNs) have been a prevailing choice. 
The cine MRI requires a reconstruction
rate of 22 frames per second (FPS) \cite{schaetz2017accelerated,iltis2015high}, setting the minimum throughput for segmentation using DNNs. In addition, the latency of the network should be less than
50 ms to avoid any lags that can be perceived by human eyes~\cite{annett2020low}.
ICA-UNet~\cite{wang2020ica} is the state-of-the-art DNN that can maintain an acceptable accuracy 
while meeting these timing constraints, with a latency of 39 ms and a throughput of 28.3 FPS.  

However, it is observed that the latency and throughput of ICA-UNet are well above the requirements. Since the MRI frames can only be reconstructed at 22 FPS, 
achieving a throughput higher than that would not yield any additional benefits but only introduce computation waste. Similarly, 
as suggested in \cite{annett2020low}, human vision cannot differentiate any latency below 50 ms. 
Therefore, there is still room to further trade off throughput and latency for higher 
accuracy. Unfortunately, ICA-UNet is manually designed, making the exploration of such tradeoff very difficult if not impossible. 

In the past a few years, differentiable neural architecture search (NAS) has been widely applied to automatically and efficiently 
identify the best neural architecture for various biomedical applications \cite{peng2021hypersegnas,zhu2021automatic,lu2021manas}. A majority of these works 
focus on improving the accuracy as the only metric of interest~\cite{he2021dints,bosma2022mixed,rguibi2021automatic}. Most recently, researchers have realized that it is crucial to also consider network latency/throughput for biomedical image segmentation applications~\cite{wang2021bix,zeng2020towards,huang2021adwu,xu2021ect,9222548}. 
They use network complexity, in terms of either the number of parameters or the number of floating point operations (FLOPs), as a proxy for latency/throughput estimation. This is primarily because latency/throughput are 
non-differentiable and can hardly be incorporated into any differentiable NAS frameworks. 
However, as demonstrated in \cite{9060902}, the latency/throughput can vary significantly even for networks of the same size and FLOPs, due to the difference in computation dependencies that 
are decided by the network 
topology. As such, there is no guarantee that the network architectures identified by these 
frameworks can satisfy the real-time constraints. As such, they cannot be extended to address
the 3D cardiac cine MRI segmentation needed for surgical visual assistance. 


To address these challenges and to directly incorporate real-time constraint into neural architecture search for 3D cardiac cine MRI segmentation, in this work, we propose a novel NAS approach, termed \textbf{RT-DNAS}, where the latency constraint can be incorporated into the differentiable search paradigm. Different from existing NAS approaches where 
both accuracy and network complexity are considered as optimization objectives, our goal is to \textit{achieve as high accuracy as possible while satisfying the latency and throughput constraints}. 
It is achieved by using a genetic algorithm (GA) to decide the final network architecture. Experiments on extended 2017 MICCAI ACDC dataset 
show that RT-DNAS can achieve the highest accuracy while satisfying the real-time constraints compared with both manual designed ICA-UNet and state-of-the-art NAS frameworks. 

\section{Method}
\label{sec:method}
\begin{figure}
    \centering
    \includegraphics[width=0.85\textwidth]{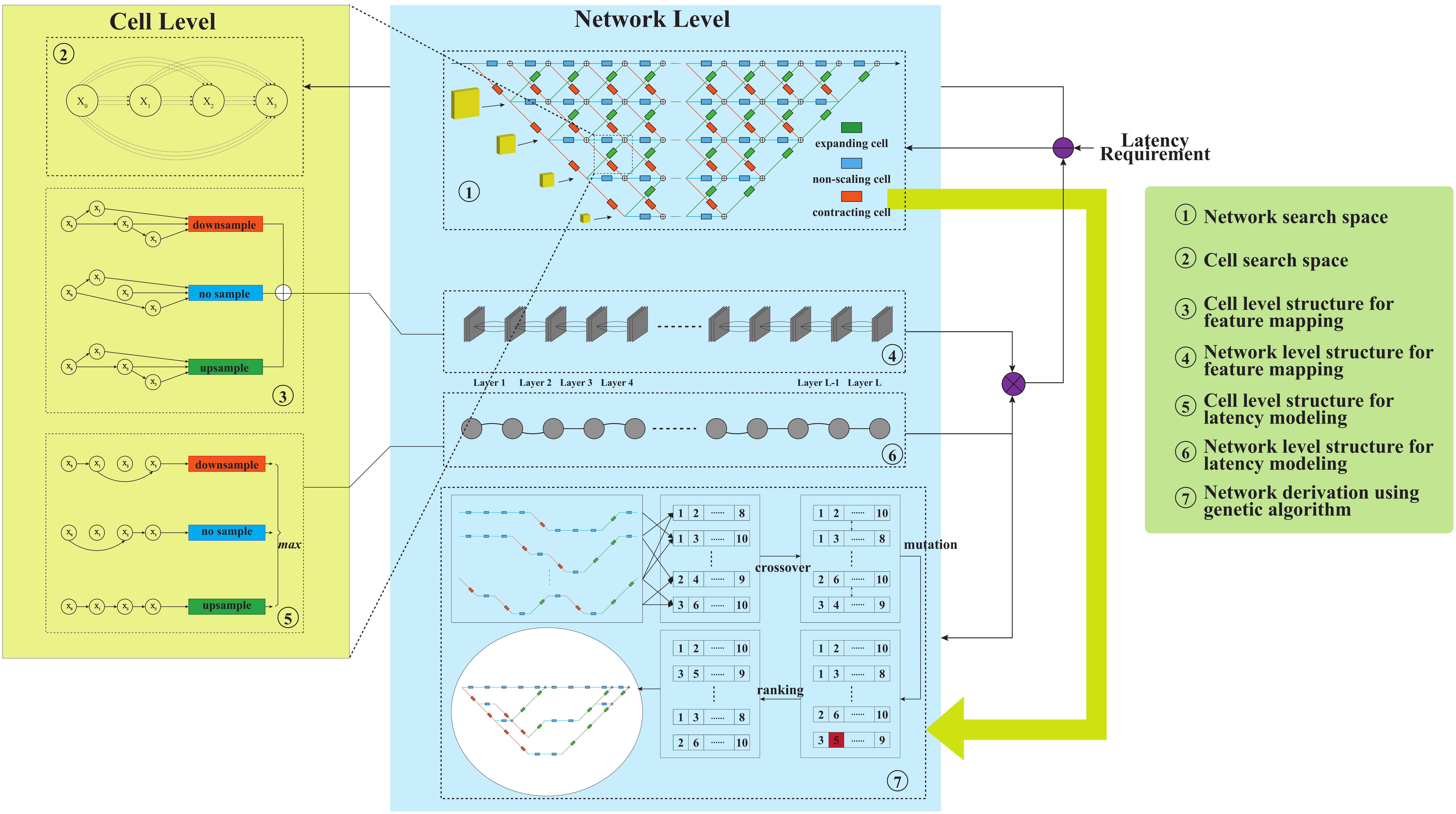}
    \caption{Overview of the RT-DNAS workflow. We adopt the MS-NAS two-level search space.
    To incorporate real-time performance, the model finds favorable cell structures within latency constraint. A modified decoding method using genetic algorithm is proposed to derive the network structure.}
    \label{fig:overview}
    
\end{figure}

We build our framework based on MS-NAS \cite{yan2020ms}, the state-of-the-art NAS approach for medical image segmentation with competitive performance and implementation flexibility. 
An overview of the framework is illustrated in Fig.~\ref{fig:overview}.

\subsection{Search Space and Method}
Inspired by MS-NAS, we define a \textit{network architecture search space} with various types of network design choices, including operation type, feature scale, feature fusion mechanism, and the connections between different operations and cells.
We define the entire NAS search space in a hierarchical manner: assume the network has $L$ layers; each layer can be composed of multiple \textit{cells}; each \textit{cell} is composed of multiple \textit{operations}.


\textbf{Cell Level}.
A cell can be represented by a directed acyclic graph (DAG) where the vertices are tensors that represent candidate operations, and the edges represent the dependence between operations. 
For an input tensor $X_0$, a cell produces $N$ new tensors $X_1$, $X_2$, ..., $X_N$, and then process all of them using a scaling operation which is the only difference among cell types.
The computation of each tensor depends on all its previous tensors (including $x_0$) and an operation $o$.
We relax the search of the connections and operations by associating an coefficient 
to each combination of input tensor, output tensor, and operation.
Denote the coefficient corresponding to the edge from $X_j$ to $X_i$ by operation $o$ as $v_{i,j}^o$, and the 
output with performing $o$ on $X_j$ is $o(X_j)$,
we can then compute $X_i$ as
\begin{equation}
    X_i = \sum_{j<i}\sum_{o\in \mathcal{O}}\frac{exp(v_{i,j}^o)}{\sum_{j'<i}\sum_{o'\in \mathcal{O}}exp(v_{i,j'}^{o'})}o(X_j)
\end{equation}
where $\mathcal{O}$ is the operation set including, depth-wise separable convolution, dilated convolution, etc.
It is worth mentioning that relaxing the connection and operation search separately as in MS-NAS can easily 
cause the gradient of relaxation coefficients vanish, so we directly search input-operation pairs instead.

\textbf{Network Level}
The network level search builds a network consisting of multiple paths from input to output; each 
paths is formed by $L$ cells in series.
This is completed by two phases.
In the first phase, all the paths are combined to generate the feature maps at all scales. 
Following the original MS-NAS, we relax the selection of different types of cells, i.e. expanding cell, non-scaling cell, and contracting cell, where their output can merge in the supernet topology (Figure. \ref{fig:overview}).
As a result, the performance each path can be evaluated by the proximal accumulated weight along the cells it covers. In the second phase, final networks are derived from the best paths based on such evaluation method. 
To take real-time efficiency into account, we propose a second search procedure using genetic algorithm to decode the network level structure as described in Section \ref{sec:arch-derivation}.

\subsection{Latency Constraint Incorporation}
Searching architectures for excellent real-time performance requires considering not only accuracy but also latency and throughput. However, since throughput is naturally coupled with latency (the higher the latency, the lower the throughput), to reduce search burden, we will incorporate latency only 
in the search process. If the identified architectures cannot satisfy the throughput constraint, we can simply lower the maximum latency and run the search again. As such, the goal of RT-DNAS is to discover the Pareto frontier in the trade-offs between the two metrics. Inspired by an early work targeting a specific hardware accelerator~\cite{0c17230688154c0c924c89ad3b81434c}, we define the loss function as follows:
\begin{equation}\label{eq:loss}
    \mathcal{L} = \mathcal{L}_a\cdot \mathcal{L}_t + \lambda e^{L_t-L_{ub}}
\end{equation}
where $\mathcal{L}_a$ is the cross-entropy loss of the network evaluated using training data, $\mathcal{L}_t$ is 
the loss for latency to ensure real-time performance; $L_t$ is the latency achieved by the network candidate; $L_{up}$ is a constant representing the latency upper bound, which is set to be 50 ms as advised by \cite{annett2020low}; $\lambda$ is a large penalty factor,
indicating that if the network architecture latency excels the upper bound, the loss will increase exponentially,
so that infeasible solutions can be eliminated as early as possible.
Since it is difficult to measure the real latency on the target device during the search, we estimate the latency $\mathcal{L}_t$ in two steps, including cell level and network level, similar to the search space construction.

\textbf{Cell level}.
For each tensor-to-tensor connection, there are $|\mathcal{O}|$ operations in parallel and the latency of the connection equals to the selected operation. 
Therefore, a reasonable estimation of \textit{latency expectation} is the weighted sum of all the latency values 
from tensor $i$ to tensor $j$, as follows:
\begin{equation}
    Lat_{i,j} = \sum_{j<i}\sum_{o\in \mathcal{O}}\frac{exp(v_{i,j}^o)}{\sum_{j'<i}\sum_{o'\in \mathcal{O}}exp(v_{i,j'}^{o'})}Lat_{i,j}^o(X_j)
\end{equation}
where $Lat_{i,j}^o$ is the latency of operation $o$ with input $X_j$, a constant for a specific operation on a fixed-size tensor.
With the latency between any pair of tensors available, the latency of the cell is computed as:
\begin{equation}\label{eq:cell-latency}
    Lat_{cell} = max_{\mathbf{x}\in\mathcal{X}}Path(\mathbf{x}) + S_{cell}
\end{equation}
where $S_{cell}$ is a constant representing the latency of scaling operation; $\mathbf{x}$ is a set of the 
tensors that form a path from $X_0$ to $X_N$; $\mathcal{X}$ is the set of all possible paths.
Note that Equation \ref{eq:cell-latency} can be solved by finding the longest path within the cell.


\textbf{Network level}.
The overall latency of a network architecture can be estimated using the summation of the latency of each layer.
Similar to cell level latency computation, we take a weighted sum of all the cell latency values to compute layer 
latency expectation based on the probability of each cell occurring in a network path as follows:
\begin{equation}
    L_t = \sum_{l\le L}\sum_{c\in \mathcal{C}^l}k_c^l{Lat_{cell}}
\end{equation}
where $k_c^l$ is the probability of cell $c$ being sampled from $l$-$th$ layer.
The details about the computation of $k_c$ are presented in the supplementary materials.

\subsection{Architecture Derivation} \label{sec:arch-derivation}
After the search, the architecture parameters $\mathcal{W}$ are decided. The final network can be derived by combining a group of paths formed by a sequence of cells from input to output. Note that different paths usually have overlapping segments in the network topology, which 
offers opportunities to further reduce the latency.
Existing work MS-NAS constructs the network by finding a fixed number of paths with highest accumulated weights to maximize the accuracy and does not consider network latency.
In fact, the weight-latency ratio of different paths varies significantly (see supplementary material),
so that carefully selecting the paths is important to reducing the network latency.

In RT-DNAS, we optimize the network path selection using genetic algorithm (GA) to build faster networks without 
significance in accuracy degradation.
We define the following concepts in GA. \textit{Gene:} a legal end-to-end path sampled from the backbone graph; \textit{Individual:} a network formed by $N_l$ paths;
\textit{Fitness:} the overall network latency of an individual.
To keep the high accuracy, a gene pool is firstly built by top paths with high accumulated weights, 
and only the selected genes can exist during the evolution. 
We apply two operations to generate new graph, crossover and mutation.
The crossover allows two individuals to exchange a portion of their genes and mutation replaces a random gene of an individual 
with another gene from the pool.
To prevent duplication, the paths are sorted by their weights and the order of genes is maintained throughout the process.

\section{Experiments}
\subsection{Experiment Setup}
For performance evaluation, following existing works, 
we employ the data of the extended 2017 MICCAI ACDC challenge dataset made available
by MSU-Net \cite{wang2019msu} with labels on all the frames in the training data. The dataset covers right ventricle (LV), myocardium (MYO), and left ventricle (LV) labeled by experienced radiologists from segmentation frames. 
There are totally 150 exams in this dataset each from a different patient, where the training set of 100 exams are employed to search the architecture.
Further, 5-fold cross validation is performed to evaluate the resultant network using Dice score.

Regarding the implementation of RT-DNAS, we follow the guidelines presented in \cite{yan2020ms}.
The skeleton is constructed by 10 layers of modules, three cells in each module, and three blocks in each cell.
$maxpooling$ and bilinear upsampling are used for contracting 
and expanding the feature size respectively, both by a factor of 2.
We also adopt partial channel connection technique \cite{xu2019pc} to reduce memory cost by forcing 3/4 channels not going through the selected operations.
We spend 40 epochs tuning the architectural parameters using an SGD optimizer with momentum of 0.9, learning rate from 0.01 to 0.001, and weight decay of 0.0003.

After the all the parameters are fixed, networks consisting of different number of paths ($N_l$) are investigated and derived using GA metohd presented in Section \ref{sec:arch-derivation}. 
The initial population is set to 20, genes are randomly selected from a pool consisting of the top 10 paths, and the search loop runs for 100 generations.
The evolution of the cell structures with the highest probabilities as well as the final architecture obtained by RT-DNAS can be visualized in the supplementary material. 


We compare RT-DNAS with ICA-UNet\cite{wang2020ica},  a manually designed network that attains state-of-the-art performance on this dataset 
while satisfying real-time 
constraints. In addition, we also apply two recent NAS frameworks to the same dataset 
with the same exploration space for fair comparison: 
hardware-aware NAS
\cite{zeng2020towards}, which includes both accuracy and FLOPs as optimization objectives, and 
BiX-NAS \cite{wang2021bix}, which includes architectural optimization. They aim at reducing 
computation complexity and parameter number, but do not explicitly take latency/throughput 
as hard constraints. All hyper-parameters of these methods are based on the values 
reported in the respective papers. 

All the networks are executed using CUDA/CuDNN~\cite{chetlur2014cudnn}. All experiments, including latency and throughput profiling, run on a machine with 16 cores of Intel Xeon E5-2620 v4 CPU, 256G memory, and an NVIDIA Tesla P100 GPU. 

\subsection{Results}
We start with an ablation study to show the benefits of deploying the GA method instead of the 
{\it Dijkstra} algorithm used in the original MS-NAS in determining the 
network level structure. 
Based on the identified cell structures by RT-DNAS, we construct sub-networks using both 
{\it Dijkstra} and GA methods, 
and the performance of the resulting networks are summarized in Table~\ref{tab:rt-dnas}. 
Comparing the two algorithms with the same number of 
consecutive paths $N_l$, 
it can be clearly observed that GA method gives higher throughput and lower latency, with slightly higher accuracy. This convincingly demonstrates the necessity and effectiveness of deploying 
GA in the network level search to handle the real-time constraints. It is also concluded that $N_l$ is an effective tuning knob in balancing between the accuracy and timing performance: the more paths involved in the sub-network, the higher Dice score can be achieved, and the slower the network runs.

\begin{table}
\vspace{-12pt}
\caption{Comparison of Dice, latency (LT, in ms) and throughput (TP, in FPS) of sub-networks formed by RT-DNAS with network level search carried out using the proposed GA algorithm (which considers real-time constraints) and the {\it Dijkstra} algorithm used in the original MS-NAS (which cannot consider real-time constraints). $N_l$ is the number of consecutive paths used.} 
\label{tab:rt-dnas}
\begin{center}
    \begin{tabular}{ccccccc}
        \hline\noalign{\smallskip}
        Methods & RV & MYO & LV & Average & LT & TP\\
        \noalign{\smallskip}
        \hline
        \noalign{\smallskip}
        $Dijkstra$ ($N_l=3$) & .865$\pm$.024 & .843$\pm$.021 & .877$\pm$.022 & .849$\pm$.032 & 27 & 39.8 \\
        $Dijkstra$ ($N_l=5$) & .911$\pm$.033 & .881$\pm$.028 & .925$\pm$.019 & .905$\pm$.020 & 40 & 32.0 \\
        $Dijkstra$ ($N_l=7$) & .928$\pm$.026 & .882$\pm$.017 & .943$\pm$.020 & .918$\pm$.032 & 48 & 22.1 \\
        \noalign{\smallskip}
        \hline
        \noalign{\smallskip}
        Genetic ($N_l=3$) & .857$\pm$.034 & .843$\pm$.020 & .878$\pm$.015 & .859$\pm$.027 & 26 & 40.4\\
        Genetic ($N_l=5$) & .910$\pm$.026 & .889$\pm$.039 & .932$\pm$.022 & .910$\pm$.031 & 35 & 34.1\\
        Genetic ($N_l=7$) & .925$\pm$.023 & .891$\pm$.019 & .950$\pm$.016 & .922$\pm$.028 & 41 & 28.9\\
        \noalign{\smallskip}
        \hline
    \end{tabular}
\end{center}
\end{table}

\begin{table}
\vspace{-24pt}
\caption{Comparison of Dice score, latency (LT, in ms) and throughput (TP, in FPS) of the state-of-the-art real-time caridac cine MRI segmentation method ICA-UNet (with two settings used 
in the original paper), and the best networks discovered by Hardware-aware NAS, BiX-NAS, and the proposed RT-DNAS. The latency cannot 
exceed 50 ms while throughput should be no less than 22 FPS. Violations of latency and throughput constraints are marked in red. 
}
\label{tab:icaunet}
\begin{center}
    \begin{tabular}{ccccccc}
        \hline\noalign{\smallskip}
        Methods & RV & MYO & LV & Average & LT & TP\\
        \noalign{\smallskip}
        \hline
        \noalign{\smallskip}
        ICA-UNet (n=3) \cite{wang2020ica} & .900$\pm$.023 & .869$\pm$.027 & .934$\pm$.013 & .901$\pm$.017 & 35 & 31.6 \\
        ICA-UNet (n=4) \cite{wang2020ica} & .921$\pm$.017 & .888$\pm$.034 & .952$\pm$.011 & .920$\pm$.019 & 39 & 28.3 \\
        HW-Aware NAS \cite{zeng2020towards} & .922$\pm$.015 & .910$\pm$.042 & .958$\pm$.019 & .930$\pm$.020 & \textcolor{red}{52} & \textcolor{red}{19.5} \\
        BiX-NAS \cite{wang2021bix} & .916$\pm$.021 &.908$\pm$.012 & .946$\pm$.033 & .923$\pm$.021 & 49 & \textcolor{red}{20.1} \\
        \hline 
        RT-DNAS ($N_l=6$) & .924$\pm$.010 & .890$\pm$.017 & .939$\pm$.030 & .918$\pm$.018 & 37 & 31.8\\
        RT-DNAS ($N_l=7$) & .925$\pm$.023 & .891$\pm$.019 & .950$\pm$.016 & .922$\pm$.028 & 41 & 28.9\\
        RT-DNAS ($N_l=8$) & \bf {.933$\pm$.018} & \bf{.902$\pm$.029} & \bf{.959$\pm$.011} & \bf{.931$\pm$.019} & 46 & 23.4\\
        \noalign{\smallskip}
        \hline
    \end{tabular}
\end{center}
\end{table}

\begin{figure}
\vspace{-12pt}
    \begin{subfigure}[b]{0.48\textwidth}
        \centering
        \includegraphics[width=\textwidth]{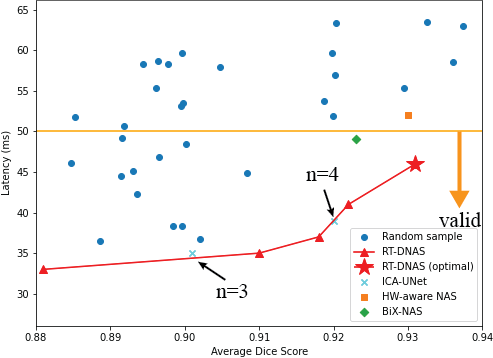}
        \caption{}
        \label{fig:dice-latency}
    \end{subfigure}
    \hfill
    \begin{subfigure}[b]{0.48\textwidth}
        \centering
        \includegraphics[width=\textwidth]{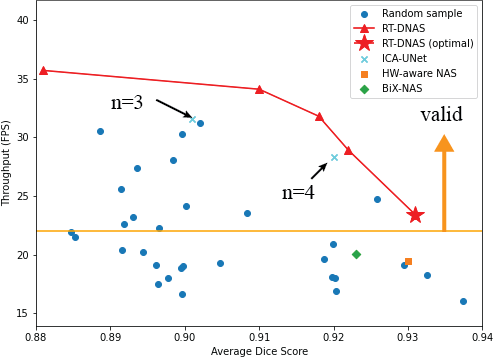}
        \caption{}
        \label{fig:dice-throughput}
    \end{subfigure}
    \caption{Accuracy v.s. (a) latency and (b) throughput for randomly sampled architectures, ICA-UNet (n=3, 4), and the best architectures identified by hardware-aware NAS, BiX-NAS, and RT-DNAS. We can see that the Pareto frontier of the points in the valid region is formed by the architectures from RT-DNAS.}
    \label{fig:my_label}
\end{figure}

We then collect the best networks obtained by RT-DNAS in terms of the tradeoff between accuracy and efficiency and compare them with ICA-UNet and the networks obtained by hardware-aware NAS and BiX-NAS. 
Note that there are two reported versions of ICA-UNet in \cite{wang2020ica}(with a hpyerparameter $n$ equal to 3 and 4 respectively). 
As shown in Table~\ref{tab:icaunet}, RT-DNAS ($N_l=6$) has slightly longer latency and similar 
throughput compared with ICA-UNet (n=3), but higher average Dice. Similar comparison can be 
observed between 
RT-DNAS ($N_l=7$) and ICA-UNet (n=4). Moreover, all these networks have a latency well below 50 ms 
and a throughput well above 22 FPS, suggesting additional room to enhance Dice. 
On the other hand, hardware-aware NAS and BiX-NAS, 
although both achieving higher average Dice, violate the real-time constraint in 
terms of either latency or throughput. This echos our discussion earlier that simply considering 
network computation complexity or parameter number cannot guarantee that the resulting networks 
will meet the real-time constraints on latency and throughput. Finally, RT-DNAS (n=8) yields 
the highest Dice in RV, MYO and LV as well as overall average among all the networks that can 
satisfy real-time constraints. Specifically, RT-DNAS (n=8) achieves 1.1\% and 3.0\% higher Dice 
compared with ICA-UNet (n=4) and (n=3), respectively. 

Finally, in Fig.~\ref{fig:my_label} we plot 
the average Dice v.s. latency/throughput for these networks as well as the randomly 
sampled architectures from the exploration space. From the figure we can see that a random
search has a very slim chance to find any close-to-optimal architectures, due to the 
large exploration space. On the other hand, the architectures identified by RT-DNAS form the 
Pareto frontier of all the points in the valid region defined by maximum latency and minimum throughput, suggesting that they can indeed offer feasible designs with best tradeoff.  
Visualization of segmentation results by RT-DNAS (n=8) along with the corresponding ground truth is included in the supplementary material.

\section{Conclusion}
In this work we proposed a latency-aware differentiable NAS framework for cine magnetic image segmentation, named RT-DNAS. 
By incorporating the hardware latency into the search objective, RT-DNAS is able to achieve the highest accuracy while perfectly meeting the real-time constraint, comparing with state-of-the-art manually and automatically designed architectures.

\bibliographystyle{splncs04}
\bibliography{bibliography}

\begin{thebibliography}{10}
\providecommand{\url}[1]{\texttt{#1}}
\providecommand{\urlprefix}{URL }
\providecommand{\doi}[1]{https://doi.org/#1}

\bibitem{annett2020low}
Annett, M., Ng, A., Dietz, P., Bischof, W.F., Gupta, A.: How low should we go?
  understanding the perception of latency while inking. In: Graphics Interface
  2014, pp. 167--174. AK Peters/CRC Press (2020)

\bibitem{bosma2022mixed}
Bosma, M., Dushatskiy, A., Grewal, M., Alderliesten, T., Bosman, P.A.:
  Mixed-block neural architecture search for medical image segmentation. SPIE
  Medical Imaging  (2022)

\bibitem{chetlur2014cudnn}
Chetlur, S., Woolley, C., Vandermersch, P., Cohen, J., Tran, J., Catanzaro, B.,
  Shelhamer, E.: cudnn: Efficient primitives for deep learning. arXiv preprint
  arXiv:1410.0759  (2014)

\bibitem{9222548}
Fernandes, F.E., Yen, G.G.: Automatic searching and pruning of deep neural
  networks for medical imaging diagnostic. IEEE Transactions on Neural Networks
  and Learning Systems  \textbf{32}(12),  5664--5674 (2021)

\bibitem{gaspar2014three}
Gaspar, T., Piorkowski, C., Gutberlet, M., Hindricks, G.: Three-dimensional
  real-time mri-guided intracardiac catheter navigation. European heart journal
   \textbf{35}(9),  589--589 (2014)

\bibitem{he2021dints}
He, Y., Yang, D., Roth, H., Zhao, C., Xu, D.: Dints: Differentiable neural
  network topology search for 3d medical image segmentation. In: Proceedings of
  the IEEE/CVF Conference on Computer Vision and Pattern Recognition. pp.
  5841--5850 (2021)

\bibitem{huang2021adwu}
Huang, Z., Wang, Z., Gu, L., et~al.: Adwu-net: Adaptive depth and width u-net
  for medical image segmentation by differentiable neural architecture search.
  Medical Imaging with Deep Learning  (2021)

\bibitem{iltis2015high}
Iltis, P.W., Frahm, J., Voit, D., Joseph, A.A., Schoonderwaldt, E.,
  Altenm{\"u}ller, E.: High-speed real-time magnetic resonance imaging of fast
  tongue movements in elite horn players. Quantitative imaging in medicine and
  surgery  \textbf{5}(3), ~374 (2015)

\bibitem{9060902}
Jiang, W., Yang, L., Sha, E.H.M., Zhuge, Q., Gu, S., Dasgupta, S., Shi, Y., Hu,
  J.: Hardware/software co-exploration of neural architectures. IEEE
  Transactions on Computer-Aided Design of Integrated Circuits and Systems
  \textbf{39}(12),  4805--4815 (2020)

\bibitem{0c17230688154c0c924c89ad3b81434c}
Li, Y., Hao, C., Zhang, X., Liu, X., Chen, Y., Xiong, J., Hwu, W., Chen, D.:
  Edd: Efficient differentiable dnn architecture and implementation co-search
  for embedded ai solutions. In: 2020 57th ACM/IEEE Design Automation
  Conference, DAC 2020. Proceedings - Design Automation Conference, Institute
  of Electrical and Electronics Engineers Inc., United States (Jul 2020)

\bibitem{lu2021manas}
Lu, Z., Xia, W., Huang, Y., Shan, H., Chen, H., Zhou, J., Zhang, Y.: Manas:
  Multi-scale and multi-level neural architecture search for low-dose ct
  denoising. arXiv preprint arXiv:2103.12995  (2021)

\bibitem{mcveigh2006real}
McVeigh, E.R., Guttman, M.A., Lederman, R.J., Li, M., Kocaturk, O., Hunt, T.,
  Kozlov, S., Horvath, K.A.: Real-time interactive mri-guided cardiac surgery:
  Aortic valve replacement using a direct apical approach. Magnetic Resonance
  in Medicine: An Official Journal of the International Society for Magnetic
  Resonance in Medicine  \textbf{56}(5),  958--964 (2006)

\bibitem{peng2021hypersegnas}
Peng, C., Myronenko, A., Hatamizadeh, A., Nath, V., Siddiquee, M.M.R., He, Y.,
  Xu, D., Chellappa, R., Yang, D.: Hypersegnas: Bridging one-shot neural
  architecture search with 3d medical image segmentation using hypernet. arXiv
  preprint arXiv:2112.10652  (2021)

\bibitem{radau2011vurtigo}
Radau, P.E., Pintilie, S., Flor, R., Biswas, L., Oduneye, S.O., Ramanan, V.,
  Anderson, K.A., Wright, G.A.: Vurtigo: visualization platform for real-time,
  mri-guided cardiac electroanatomic mapping. In: International Workshop on
  Statistical Atlases and Computational Models of the Heart. pp. 244--253.
  Springer (2011)

\bibitem{rguibi2021automatic}
Rguibi, Z., Hajami, A., Dya, Z.: Automatic searching of deep neural networks
  for medical imaging diagnostic. In: International Conference on Advanced
  Technologies for Humanity. pp. 129--140. Springer (2021)

\bibitem{rogers2016transcatheter}
Rogers, T., Mahapatra, S., Kim, S., Eckhaus, M.A., Schenke, W.H., Mazal, J.R.,
  Campbell-Washburn, A., Sonmez, M., Faranesh, A.Z., Ratnayaka, K., et~al.:
  Transcatheter myocardial needle chemoablation during real-time magnetic
  resonance imaging: a new approach to ablation therapy for rhythm disorders.
  Circulation: Arrhythmia and Electrophysiology  \textbf{9}(4),  e003926 (2016)

\bibitem{schaetz2017accelerated}
Schaetz, S., Voit, D., Frahm, J., Uecker, M.: Accelerated computing in magnetic
  resonance imaging: Real-time imaging using nonlinear inverse reconstruction.
  Computational and mathematical methods in medicine  \textbf{2017} (2017)

\bibitem{vergara2011real}
Vergara, G.R., Vijayakumar, S., Kholmovski, E.G., Blauer, J.J., Guttman, M.A.,
  Gloschat, C., Payne, G., Vij, K., Akoum, N.W., Daccarett, M., et~al.:
  Real-time magnetic resonance imaging--guided radiofrequency atrial ablation
  and visualization of lesion formation at 3 tesla. Heart Rhythm
  \textbf{8}(2),  295--303 (2011)

\bibitem{wang2019msu}
Wang, T., Xiong, J., Xu, X., Jiang, M., Yuan, H., Huang, M., Zhuang, J., Shi,
  Y.: Msu-net: Multiscale statistical u-net for real-time 3d cardiac mri video
  segmentation. In: International Conference on Medical Image Computing and
  Computer-Assisted Intervention. pp. 614--622. Springer (2019)

\bibitem{wang2020ica}
Wang, T., Xu, X., Xiong, J., Jia, Q., Yuan, H., Huang, M., Zhuang, J., Shi, Y.:
  Ica-unet: Ica inspired statistical unet for real-time 3d cardiac cine mri
  segmentation. In: International conference on medical image computing and
  computer-assisted intervention. pp. 447--457. Springer (2020)

\bibitem{wang2021bix}
Wang, X., Xiang, T., Zhang, C., Song, Y., Liu, D., Huang, H., Cai, W.: Bix-nas:
  Searching efficient bi-directional architecture for medical image
  segmentation. In: International Conference on Medical Image Computing and
  Computer-Assisted Intervention. pp. 229--238. Springer (2021)

\bibitem{xu2021ect}
Xu, S., Quan, H.: Ect-nas: Searching efficient cnn-transformers architecture
  for medical image segmentation. In: 2021 IEEE International Conference on
  Bioinformatics and Biomedicine (BIBM). pp. 1601--1604. IEEE (2021)

\bibitem{xu2019pc}
Xu, Y., Xie, L., Zhang, X., Chen, X., Qi, G.J., Tian, Q., Xiong, H.: Pc-darts:
  Partial channel connections for memory-efficient architecture search. arXiv
  preprint arXiv:1907.05737  (2019)

\bibitem{yan2020ms}
Yan, X., Jiang, W., Shi, Y., Zhuo, C.: Ms-nas: Multi-scale neural architecture
  search for medical image segmentation. In: International Conference on
  Medical Image Computing and Computer-Assisted Intervention. pp. 388--397.
  Springer (2020)

\bibitem{zeng2020towards}
Zeng, D., Jiang, W., Wang, T., Xu, X., Yuan, H., Huang, M., Zhuang, J., Hu, J.,
  Shi, Y.: Towards cardiac intervention assistance: hardware-aware neural
  architecture exploration for real-time 3d cardiac cine mri segmentation. In:
  Proceedings of the 39th International Conference on Computer-Aided Design.
  pp.~1--8 (2020)

\bibitem{zhu2021automatic}
Zhu, Y., Meijering, E.: Automatic improvement of deep learning-based cell
  segmentation in time-lapse microscopy by neural architecture search.
  Bioinformatics  \textbf{37}(24),  4844--4850 (2021)

\end{thebibliography}

\end{document}


\title{Supplementary Material}
\author{}
\institute{}
%
%

%
\maketitle  
\begin{figure}
    \centering
    \includegraphics[width=0.8\textwidth]{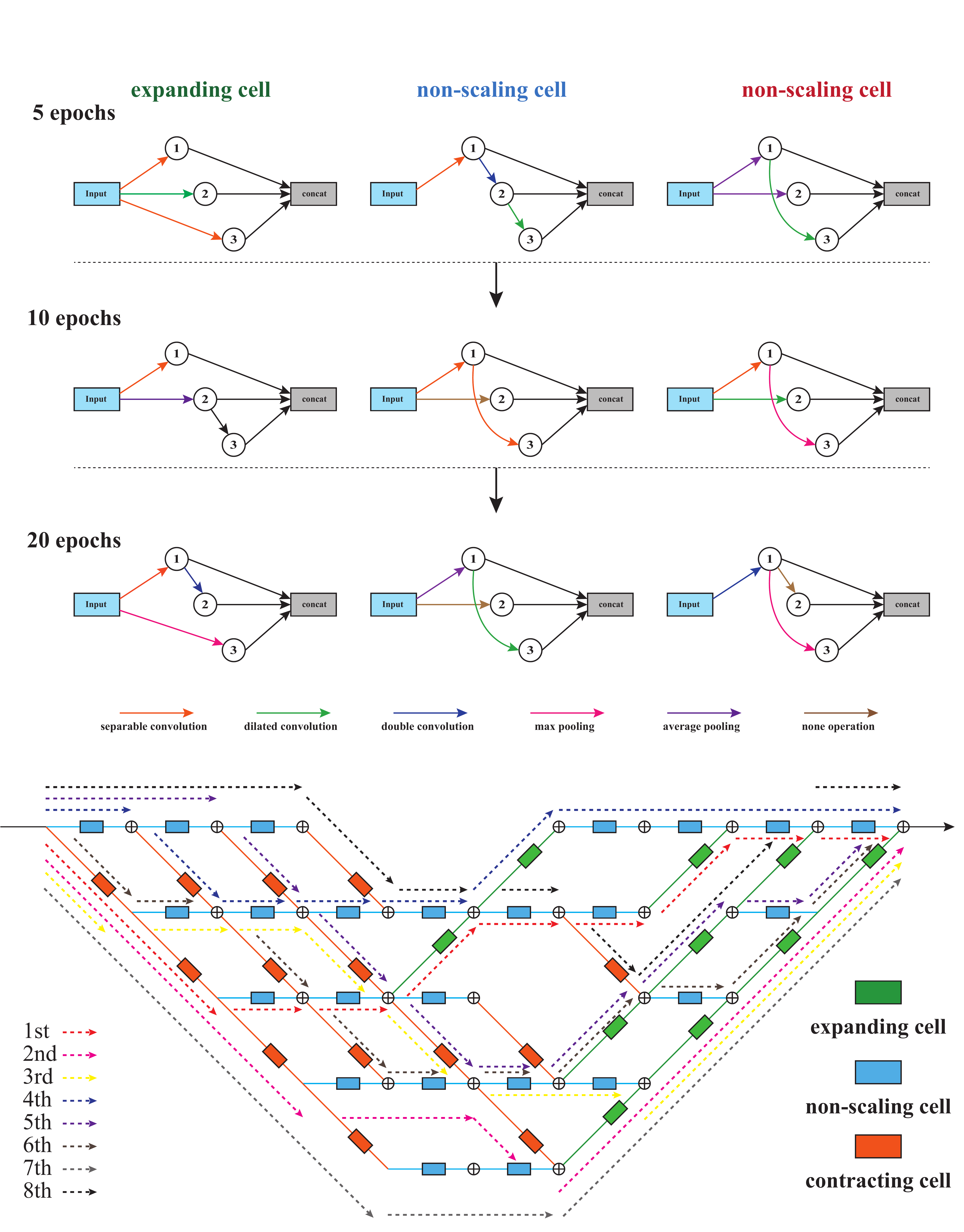}
    \caption{Illustration of the evolution of cell structure and the optimized network architecture with the best accuracy satisfying
    timing constraint discovered by RT-DNAS ($N_l=8$).}
    \label{fig:architecture}
\end{figure}

\begin{figure}
    \centering
    \includegraphics[width=0.8\textwidth]{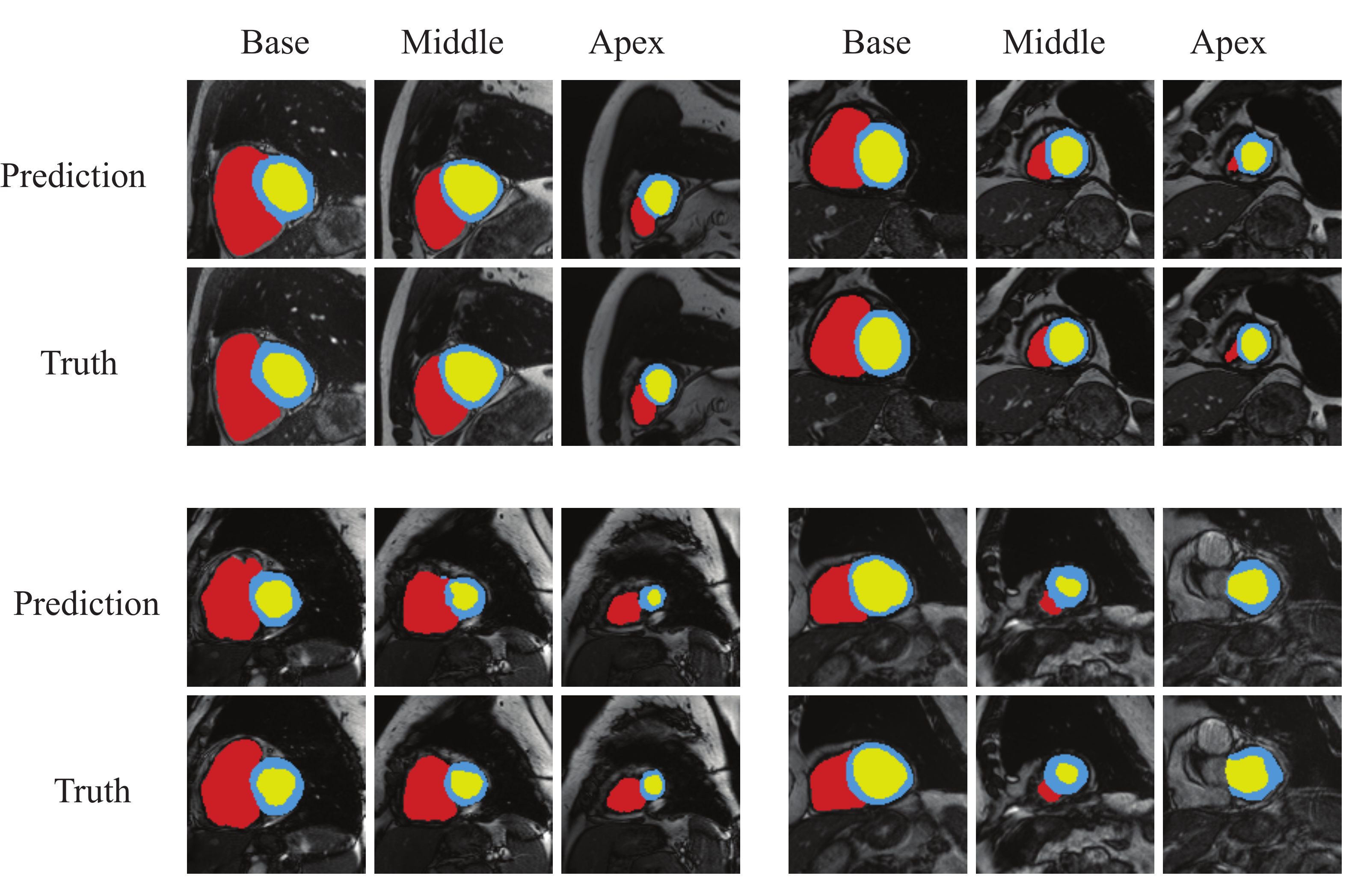}
    \caption{Visualization of the segmentation result by RT-DNAS ($N_l=8$) on cine MRI images. Four cases of different patients 
    are presented. Three different positions of LV including base, middle, apex in the same image are laid out side by side. 
    The labels of RV, MYO, and LV are marked in red, blue, and yellow, respectively.}
    \label{fig:segmentation}
\end{figure}

\begin{figure}
    \centering
    \includegraphics[width=0.7\textwidth]{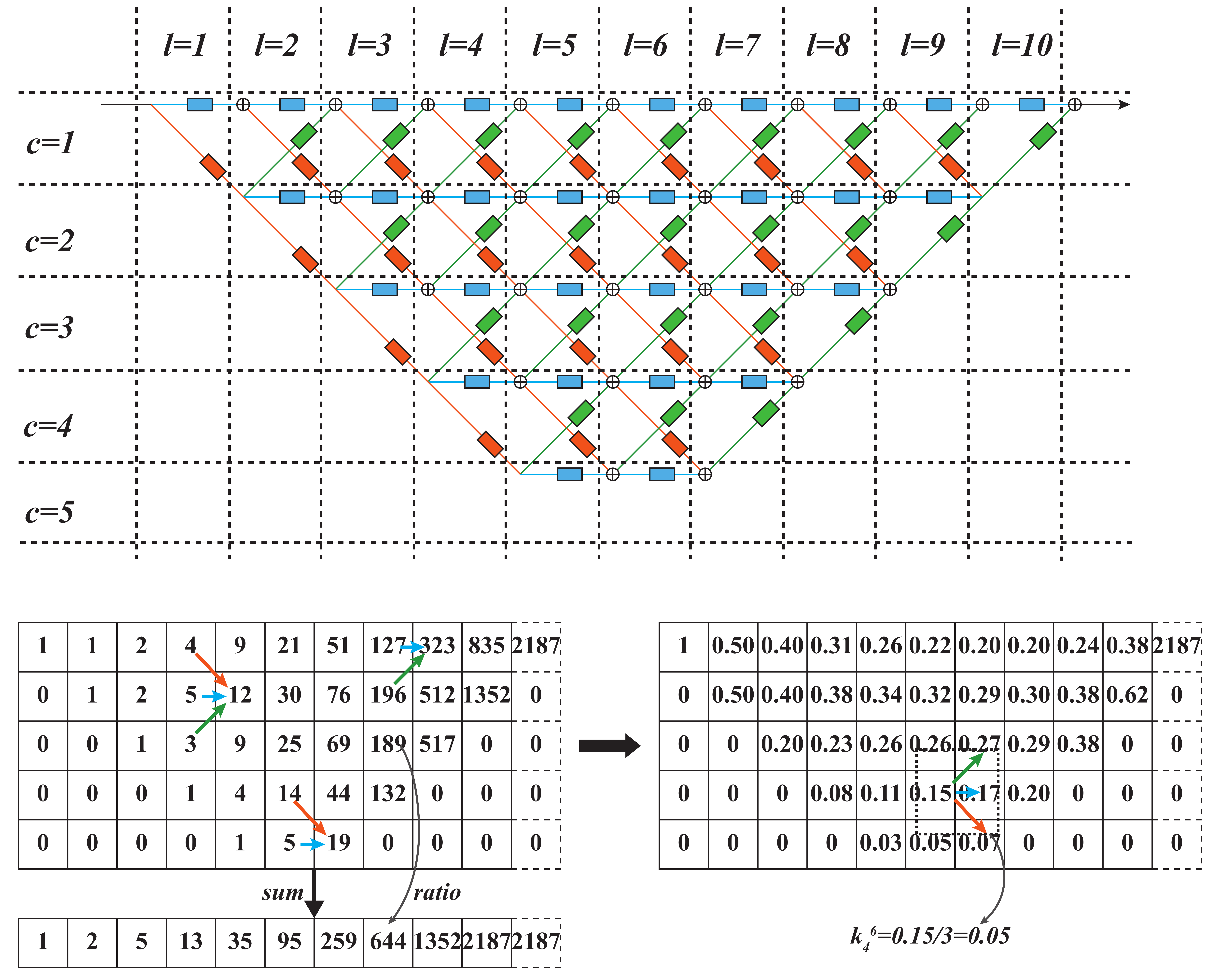}
    \caption{Computation of $k_c^l$ based on the probability with each cell occurring in a path. The supernet can be 
    divided by a grid with each cell connecting two tiles in consecutive layers. The total number of paths arriving at each tile can be explicitly 
    calculated so the probability of its occurrence is derived by the path number ratio in that layer. Trainable coefficients are not considered in network level latency model.}
    \label{fig:my_label}
\end{figure}